\documentclass{aa}
\usepackage{graphicx}
\begin{document}
   \title{Modeling the gas reservoir of circumstellar disks around young G-type stars}
   \titlerunning{Modeling gas reservoir around young G-type stars}

   \author{Inga Kamp
          \inst{1,2}
          \and
          Fatima Sammar\inst{1}
          }
   \authorrunning{Kamp \& Sammar}

   \offprints{Inga Kamp}

   \institute{Leiden Observatory, PO Box 9513, 2300 RA Leiden, The Netherlands\\
              \and Space Telescope Science Institute, 3700 San Martin Drive, Baltimore,
              MD 21218, USA, e-mail: {\tt kamp@stsci.edu}
             }

   \date{Received 27.8.2003; 27.7.2004}

   \abstract{Interpretation of molecular line observations in tenuous
             circumstellar disks around young G-type stars in terms of a
             disk mass is difficult without a model that describes the
             chemical structure of these disks. This paper now discusses
             the chemistry in tenuous disks around young solar-type stars 
             based on disk
             models that take into account the presence of a stellar
             chromosphere. The example of the disk around a 70~Myr old solar-type
             star shows that the dissociating radiation from the chromosphere is 
             stronger than the interstellar ultraviolet radiation field
             up to a distance of $\sim$~400 AU from the star. Similar to 
             other studies in this research field, it is found that, due 
             to photodissociation, the CO-to-H$_2$ ratio is far from the
             canonical value of $10^{-4}$ for molecular clouds. Moreover,
             the dust-to-gas mass ratio, as well as the dust grain size 
             play an important role for the H$_2$ abundance in these
             disks. 
   }

   \maketitle

\section{Introduction}

The IRAS and ISO satellites revealed that a large fraction of
nearby pre-main sequence solar-type stars are surrounded by cool 
dust. Near-infrared and submillimeter observations have shown 
that the dust is distributed in disk-like sometimes ring-like 
structures around the star (see Zuckerman \cite{Zuckerman:2001} 
for a recent review). However, the gas content of these 
protoplanetary disks is still under debate. CO observations by
Zuckerman et al. (\cite{Zuckerman_etal:1995}) for a sample of 
young stars in the range $10^6$-$10^7$~yr lead to the conclusion
that the gas dissipates very rapidly --- within a few million
years --- from the protoplanetary disks. A later study by
Coulson et al. (\cite{Coulson_etal:1998}) for a sample of
24 candidate Vega-excess stars shows also too little or no CO
emission. Only recently, Greaves et al. (\cite{Greaves_etal:2000}) 
found no evidence for CO rotational lines in a sample of 14 nearby 
F and G-type stars with known circumstellar dust. Since solar-type
stars are thought to have little UV radiation, the non-detection
of CO is generally interpreted as an overall lack of gas in the 
disks around these stars. As a consequence planets would have
to form either faster in the framework of the core-accretion model 
or they form on a very short timescale --- a few thousand years --- 
via gravitational instabilities.

What people have neglected so far, is the possible existence of
a chromosphere in solar-type stars. Ayres (\cite{Ayres:1997})
has shown that the ionizing ultraviolet flux from our Sun was 
much stronger in the past and evolved roughly as $t^{-1}$ with time. 
The aim of this paper is to study the influence of a chromosphere on
the chemistry in late phases of disk evolution, that is for
disks with $L_{\rm IR}/L_\ast < 0.01$. For this purpose, we
choose a template for a 70 Myr old solar-type star and compile a 
typical chromospheric radiation field (Sect.~2). The modeling 
procedure is briefly described
in Sect.~3 and the outcome of the model calculations is presented and
discussed in Sect.~4.

\section{The template G-type star}

The initial search for a template star was lead by the following criteria:
\begin{itemize}
\item The spectral type has to be similar to our Sun, G5{\sc V}.
\item $L_{\rm IR}/L_\ast < 0.01$, that is the disk has to be optically
      thin in the UV.
\item Disk parameters have to be known from infrared photometry.
\item CO observations have to exist (either detection or upper limit).
\end{itemize}
From a literature study of the papers by Sylvester \& Skinner
(\cite{SylvesterSkinner:1996}), Sylvester et al. (\cite{Sylvester_etal:1996}), 
Dunkin et al. (\cite{Dunkin_etal:1997a}, \cite{Dunkin_etal:1997b}), Zuckerman et al. 
(\cite{Zuckerman_etal:1995}), Coulson et al. (\cite{Coulson_etal:1998}) 
and Greaves et al. (\cite{Greaves_etal:2000}), only one candidate star was found:
HD\,123160. Unfortunately, the spectral type as well as the nature of
the infrared excess of this star was recently questioned: 
Mora et al. (\cite{Mora_etal:2001}) and Kalas et al. (\cite{Kalas_etal:2002}) 
classify this star as a K5 giant and detect a complex reflection nebula instead of a
disk-like structure. We conclude that observational data is simply up to now too 
scarce, and we proceed by compiling a template star using typical stellar and 
disk properties.

\subsection{Stellar parameters and disk properties}

The basic stellar parameters and disk parameters assumed
for the template star are summarized in Table~\ref{tab:HD123160}.
We assume in the following a typical age of 70~Myr for our template.

   \begin{table}
      \caption[]{Stellar parameters and disk parameters for the template star.
                 The disk parameters in brackets refer to an alternative
                 disk model
                 that would result in a very similar IR excess.}
         \label{tab:HD123160}
   \begin{tabular}{lll}
   parameter             &  template star  \\[2mm]
   \hline\\
   spectral type         &  G5\,{\sc V}       \\
   $T_{\rm eff}$         &  5570              \\
   $\log g$              &  4.5               \\
   stellar radius        &  0.9~$R_\odot$     \\
   distance              &  15.7~pc           \\
   age                   &  $70\,10^6$ yr     \\[2mm]
   $R_{\rm i}$ (disk)    &  190~AU            \\
                         &  (60~AU)           \\
   $R_{\rm o}$ (disk)    &  470~AU            \\
   $L_{\rm IR}/L_\ast$   &  $4.4\,10^{-3}$    \\
   $M_{\rm dust}$ (disk) &  0.057~$M_\oplus$  \\
                         &  (0.77~$M_\oplus$) \\
   \hline\\
   \end{tabular}
   \end{table}

Given the fact that {observations yield in most cases only a lower limit for} 
the gas mass, the disk models are calculated for three different gas masses: 
0.033, 0.33 and 3.3~M$_\oplus$. On the other hand, the dust mass is fixed to 
0.057~M$_\oplus$. This means that the gas-to-dust mass ratio varies in the 
three models from 0.58 to 58. In addition, model calculations are performed
with an alternative more massive dust disk model, M$_{\rm dust} = 0.77$~M$_\oplus$,
which shows a very similar IR excess. 

\subsection{The UV radiation field}
\label{Sect:UVradiationfield}

Following Table~1, a Kurucz ATLAS9 stellar atmosphere model 
(Kurucz \cite{Kurucz:1992}) with an effective temperature of 5750~K and a 
$\log g$ of 4.5 is assumed for our template star.

In early stages of star formation, the ultraviolet radiation field is
dominated by active accretion from the circumstellar disk/envelope. In
later stages, when accretion is no longer the dominant process, stellar
activity takes over. Studies of open star clusters have shown that
young solar-type stars are more active than the present Sun. 
Ayres (\cite{Ayres:1997}) estimated correlations between stellar activity 
and rotation for different tracers like \ion{O}{vi}, \ion{C}{iii} and Ly\,$\alpha$
based on observed X-ray-rotation and \ion{C}{iv}-rotation relations. 
Combining this with observed age-rotation relations, Ayres concludes
that all the photorates scale approximately as $\sim t^{-1}$ with time.
Photodissociation of CO and H$_2$ takes place through discrete bands and 
is thus affected mostly by the continuum. Nevertheless, some of the
dissociating bands overlap with the emission lines of an active solar-type
star, e.g.\ with the \ion{O}{vi} emission around 1038~\AA\  or the
\ion{C}{iii} at 977~\AA . Hence, it is assumed that the total flux in the
912 - 1110 \AA\  region scales with time as $\sim t^{-1}$.

   \begin{figure}[htb]
   \centering
   \includegraphics{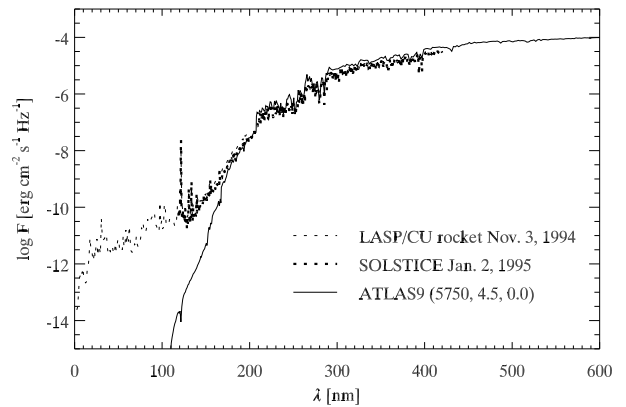}
      \caption{Flux distribution of the Sun at the solar surface. Solid line:
               Kurucz ATLAS9 model with T$_{\rm eff}=5750$~K and $\log g = 4.5$.
               Thin dotted line: LASP/CU rocket experiment. Thick dotted
               line: SOLSTICE.
              }
         \label{Sun_UV}
   \end{figure}

Data from the LASP/CU rocket experiment and Solar Stellar Irradiance Comparison 
Experiment (SOLSTICE) serves as a template for the UV radiation of our present Sun.
Fig.~\ref{Sun_UV} shows that the rocket and satellite data agrees well with
a Kurucz ATLAS9 model of 5750~K and a $\log g$ of 4.5.

   \begin{figure}[htb]
   \centering
   \includegraphics{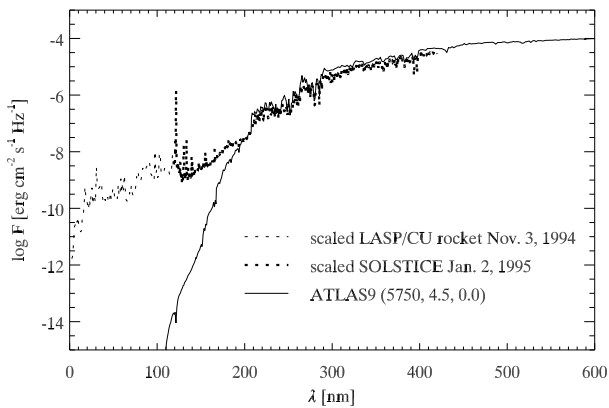}
      \caption{Flux distribution at the stellar surface of the template star. Solid line:
               Kurucz ATLAS9 model with T$_{\rm eff}=5750$~K and $\log g = 4.5$.
               Thin dotted line: LASP/CU rocket experiment data scaled by a factor 70. 
               Thick dotted line: SOLSTICE data scaled according to Eq.(\ref{scale}).
              }
         \label{HD123160_UV}
   \end{figure}

With an age of 70~Myr for our template star, the flux has to be scaled by 
a factor $\alpha =70$ with regard to the Sun. Since Ayres (\cite{Ayres:1997}) 
states that the linear scaling 
of the flux does not hold for processes involving the continuum at wavelength 
larger than 1500~\AA, a different scaling law is used for the 912 - 1190~\AA\
range (rocket data) and the 1190 - 2000~\AA\ range (SOLSTICE data). The first
wavelength interval is scaled up by a constant factor of 70. To match
the photospheric flux at 2000~\AA , the latter is then scaled according
to the following relation
\begin{eqnarray}
\log F_\nu ({\rm template}) & = & \log F_\nu({\rm SOLSTICE}) + \nonumber \\ 
                            &   & \frac{\log \alpha}{81} (2000 - \lambda)\,\,\,,
\label{scale}
\end{eqnarray}
where $F_\nu$ is given in erg~cm$^{-2}$~s$^{-1}$~Hz$^{-1}$ and $\lambda$ is in \AA .
The resulting flux distribution is depicted in Fig.~\ref{HD123160_UV}. Our high
resolution spectrum overlaps with previously published reconstructed near UV 
fluxes for the 70 Myr old Sun (Dorren \& Guinan \cite{DorrenGuinan:94}). The latter 
does not cover the UV region of interest, where CO and H$_2$ photodissociate. However, 
in the overlap region beyond 1250~\AA , the overall agreement is very good.

\begin{figure}[htb]
   \centering
   \includegraphics{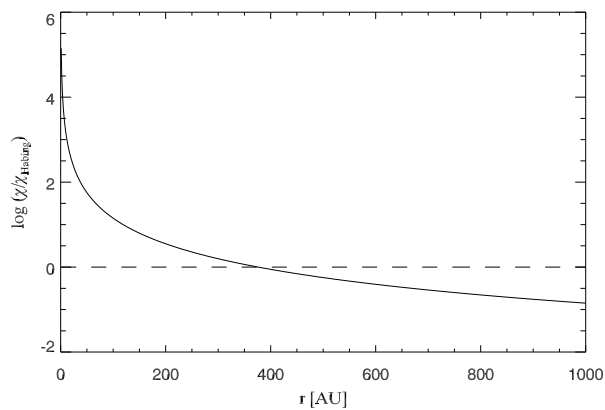}
      \caption{Unattenuated integrated UV flux (912 - 1110~\AA ) normalized
               to the interstellar Habing flux as a function of distance from
               the star: solid line template star plus chromosphere, dashed line:
               Habing field.
              }
         \label{HD123160_chi}
   \end{figure}

Fig.~\ref{HD123160_chi} illustrates the strength $\chi$ of the scaled UV radiation
field with respect to an interstellar Habing field (Habing \cite{Habing:1968}),
$1.222\,10^7$~cm$^{-2}$~s$^{-1}$. $\chi$ denotes the integrated UV flux 
over the range 912 - 1110~\AA . It is clearly seen that the stellar radiation field dominates
up to $\sim 400$~AU in an optically thin disk.

\section{The disk model}

   \begin{figure*}
   \includegraphics[width=18cm]{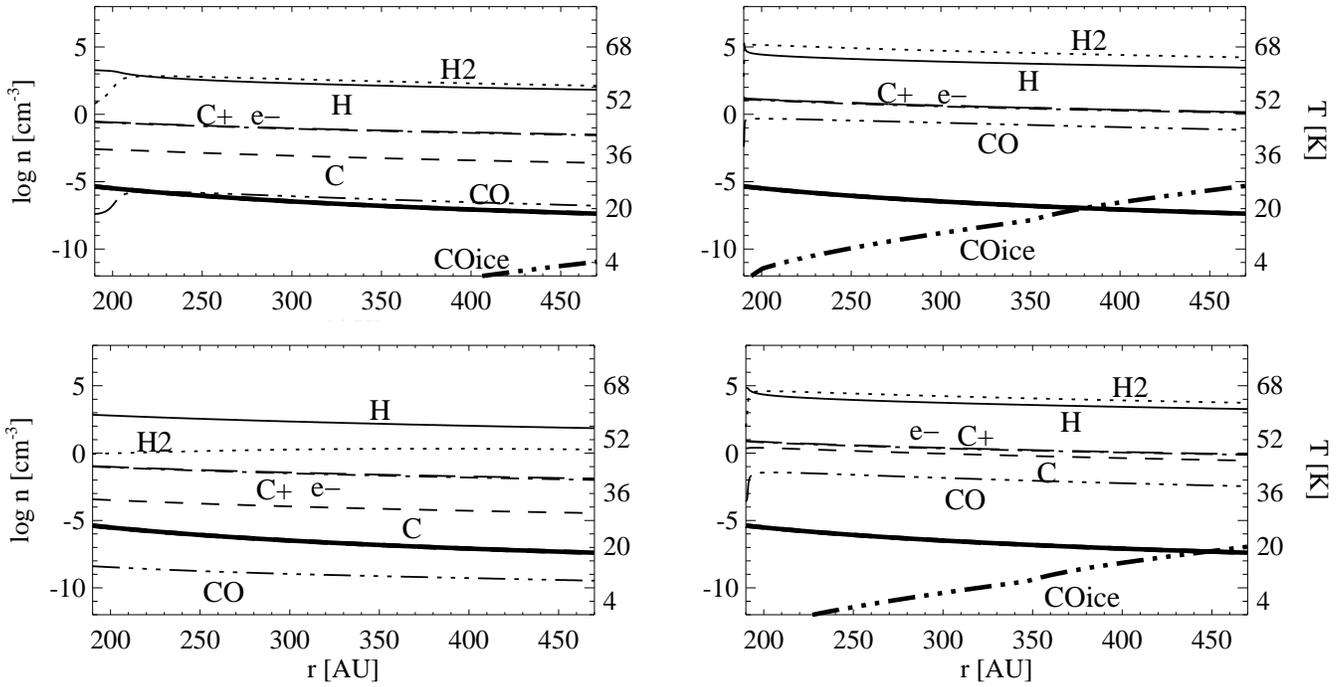}
   \caption{Densities of different species (left scale) and 
            temperature (right scale) in 2 different 
            disk models:
            (a) m0.033\_190\_CH model at the disk midplane, 
            (b) m3.3\_190\_CH model at the disk midplane, 
            (c) m0.033\_190\_CH model at one scaleheight,
            (d) m3.3\_190\_CH model at one scaleheight. The
            adopted linestyles are: H:solid, H$_2$: dotted,
            C: dashed, C$^+$: dash-dotted, CO: dash-dot-dotted,
            CO$_{\rm ice}$: dash-dot-dotted (thick), e$^-$:
            long dashed, $T$: solid (thick).}
    \label{m0.033_m3.3_chrom_chem}
    \end{figure*}

The main features of the disk models are briefly summarized and for
further details, we refer to the original paper Kamp \& Bertoldi 
(\cite{KampBertoldi:2000}).

We assume thin hydrostatic equilibrium models
\begin{equation}
        n(r,z) ~=~n_{\rm i} ~(r/R_{\rm i})^{-2.5}~e^{-z^2/2h^2}\,\,\,.
        \label{eq:density}
\end{equation}
As a simplification, we assume a dimensionless scale-height 
$H \equiv h/r = 0.15$. This does not lead to fully
self-consistent disk models, but a coupling of the gas temperature
calculation and the disk structure calculation is beyond the scope
of this paper. The inner and
outer radius of the disk are set to the respective values given in 
Table~\ref{tab:HD123160}. The power-law exponent of the disk surface 
density is $-1.5$. The stellar radiation field $F_\nu$ is described 
above (see Sect.~\ref{Sect:UVradiationfield}).

The dust temperature follows from radiative equilibrium assuming 
large spherical grains of radius $a=3~\mu$m (Kamp \& Bertoldi 
\cite{KampBertoldi:2000})
\begin{equation}
T_{\rm dust} = 282.5~\left(L_{\ast}/L_\odot\right)^{1/5} 
                             \left(r/{\rm AU}\right)^{-2/5} 
                             \left(a/{\rm\mu m}\right)^{-1/5}~,
\end{equation}
with the stellar luminosity in units of the solar luminosity
$L_\odot$. The assumption of this simple form of radiative 
equilibrium is certainly correct for the large dust grains and 
optically thin disks described in this paper.

The chemical network consists of 47 atomic, ionic, and molecular
species that are related through 268 gas-phase chemical and
photoreactions. A number of reactions is treated in more detail like
H$_2$ and CO photo-dissociation, and C ionization. The only surface
reactions incorporated are H$_2$ formation and freezing out of CO on
cold dust-grain surfaces.  Since we are dealing with large dust
particles, we reduced the H$_2$ formation rate according to the
reduced grain surface area. The abundance of CO ice is due to a
balance between freezing out of gaseous CO and reevaporation of CO
ice. The two differences to the Kamp \& Bertoldi (\cite{KampBertoldi:2000})
paper are: (1) the inclusion of cosmic ray reactions and (2)
a lower temperature of 20~K for freezing out of CO ice. However, due
to the strong stellar UV radiation field, the chemistry is mostly
driven by stellar photons. A modified Newton-Raphson algorithm is 
used to obtain a stationary solution of the entire chemical network.

\section{Results}

Four main questions are adressed in the following subsections: 
\begin{itemize}
\item What is the chemical composition of the disk? How does it
      depend on disk mass, what is the role of shielding?
\item How does the CO-to-H$_2$ ratio change with disk mass?
\item How does the varying dust-to-gas mass ratio affect the
      chemical composition of the disk?
\item What is the effect of a chromosphere as compared to the
      interstellar radiation field? When does a chromosphere
      need to be included in disk modeling?
\end{itemize}
The basic parameters of the disk models computed to answer
these questions are summarized in Table~\ref{tab:diskmodels}.
The outer radius is fixed to 470~AU in all models. The name
of the disk model is composed as follows: 
$M_{\rm gas}$\_$R_{\rm i}$\_$F_\nu$. The type of radiation field
$F_\nu$ is indicated by 'CH' or 'IS', which stands for
'chromosphere' or 'interstellar'. The former radiation field
is explained to great detail in Sect.~\ref{Sect:UVradiationfield}.
The interstellar radiation field is taken from Habing (\cite{Habing:1968}) and
approximated by the following expression
\begin{eqnarray}
F_\nu & = & \frac{10^{-14}}{f_5^2} \left( -\frac{25}{6}f_5^{-2} + \frac{25}{2}f_5^{-1} 
                                      - \frac{13}{3} \right) \nonumber \\
      &   &                         {\rm erg~cm^{-2}~s^{-1}~Hz^{-1}}~,
\end{eqnarray}
with $f_5$ being the wavenumber in $10^5$~cm$^{-1}$. For simplicity,
it is assumed that the interstellar radiation field only penetrates
the disk from the inner edge. A more realistic approach would
involve a 2D radiative transfer, which is so far not included 
in the model. 

   \begin{table}
      \caption[]{Model parameters for the various disk models. The columns denote
                 the inner and outer radius in AU, the gas mass in M$_{\oplus}$, 
                 the dust-to-gas mass ratio $\delta$, the dust absorption cross section in the 
                 UV in cm$^2$ per H-atom and the type of
                 UV radiation field. In the last colum, 'CH' denotes a pure
                 chromospheric UV radiation field, while 'IS' stands for an
                 interstellar Habing radiation field with $G_0=1$.}
         \label{tab:diskmodels}
   \begin{tabular}{lrlllc}
   name             &  $R_{\rm i}$   & $M_{\rm gas}$ & 
                       $\delta$ & $\sigma$ & $F_\nu$ \\[2mm]
   \hline\\
   m0.033\_190\_CH & 190 & 0.033 &  1.73     & 3.37(-21) & CH \\
   m0.33\_190\_CH  & 190 & 0.33  &  1.73(-1) & 3.37(-22) & CH \\
   m3.3\_190\_CH   & 190 & 3.3   &  1.73(-2) & 3.37(-23) & CH \\
   m0.033\_190\_IS & 190 & 0.033 &  1.73     & 3.37(-21) & IS \\
   m0.33\_190\_IS  & 190 & 0.33  &  1.73(-1) & 3.37(-22) & IS \\
   m3.3\_190\_IS   & 190 & 3.3   &  1.73(-2) & 3.37(-23) & IS \\
   m0.033\_60\_CH  &  60 & 0.033 &  2.3(1)   & 5.38(-20) & CH \\
   m3.3\_60\_CH    &  60 & 3.3   &  2.3(-1)  & 5.38(-22) & CH \\
   m0.033\_60\_IS  &  60 & 0.033 &  2.3(1)   & 5.38(-20) & IS \\
   m3.3\_60\_IS    &  60 & 3.3   &  2.3(-1)  & 5.38(-22) & IS \\
   \hline\\
   \end{tabular}

   Note: numbers in parentheses: 3.37(-21) = $3.37\,10^{-21}$
   \end{table}

   \begin{figure*}
   \includegraphics{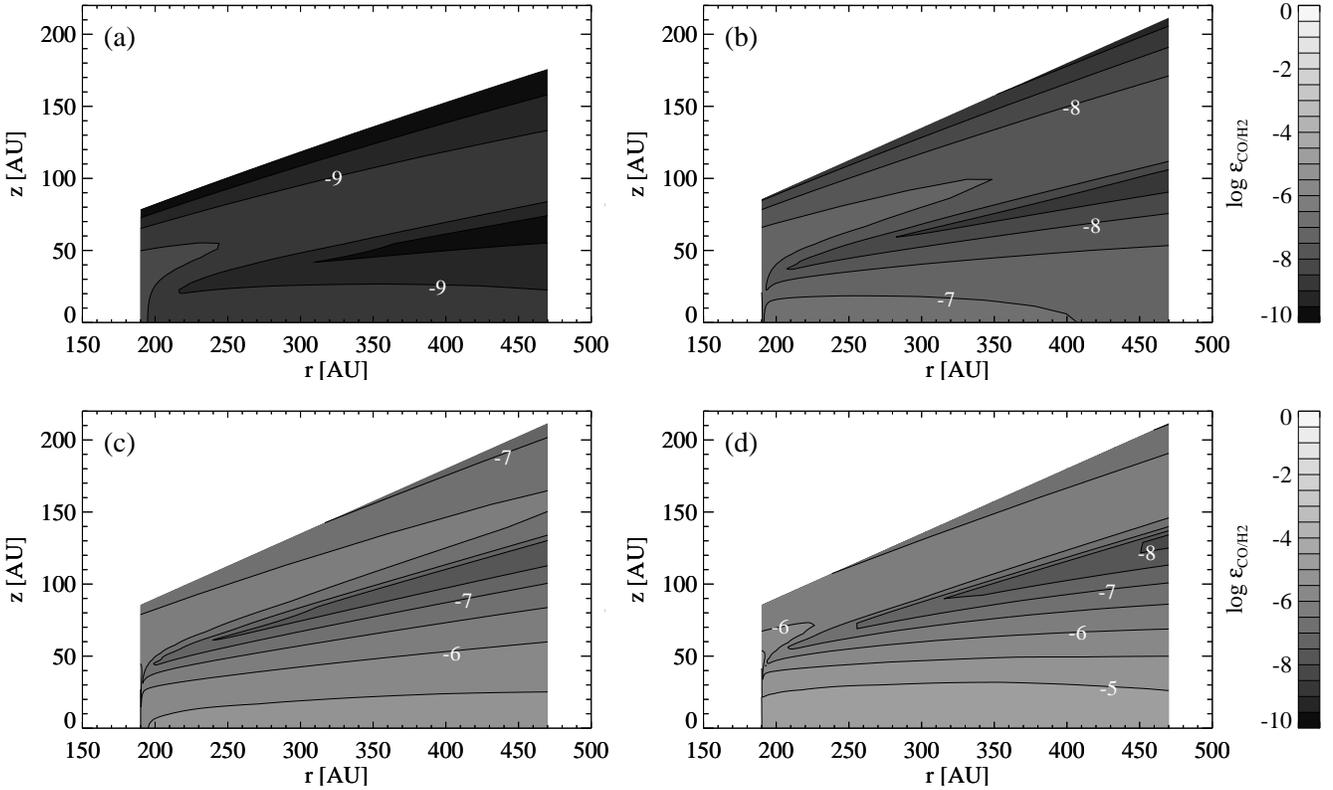}
   \caption{The CO-to-H$_2$ ratio $\epsilon$ in 4 different 
            disk models:
            (a) m0.033\_190\_CH, (b) m0.33\_190\_CH, (c)
            m3.3\_190\_CH, and (d) m3.3\_190\_IS.}
    \label{CO-to-H2}
    \end{figure*}

\subsection{The chemical composition of the disk}

We limit the discussion here to some key species, namely
H, H$_2$, C, C$^+$, CO, and CO ice, that can act as possible
tracers for the disk mass. The three models m0.033\_190\_CH,
m0.33\_190\_CH, and m3.3\_190\_CH cover the transition
from a completely transparent disk to a moderately shielded
one.

Fig.~\ref{m0.033_m3.3_chrom_chem} illustrates the chemical
structure of two disk models: m0.033\_190\_CH and m3.3\_190\_CH.
Since the disks are optically thin in the UV, dust shielding 
is not important for the photoreactions. H$_2$ can efficiently
shield itself against photodissociation within one scaleheight
in all three models. In the m3.3\_190\_CH model the molecular
hydrogen extends even to 1.5 scaleheights. Despite this high
abundance of molecular hydrogen, H$_2$ is not able to shield 
CO against photodissociation. Self-shielding of CO sets in
for the m0.33\_190\_CH disk model and becomes efficient in
the highest mass model, m3.3\_190\_CH. There, the CO abundance
reaches values of $\log \epsilon_{\rm CO}=-5.4$. In the outermost 
parts of the disk, $r > 390$~AU, the temperature drops below
20~K, but the low densities $n_{\rm tot}<10^5$~cm$^{-3}$ 
prevent a strong freeze-out of CO on dust grains.

The chemistry depends very much on the He abundance: Helium is 
partially ionized by cosmic rays. The ionization degree 
\ion{He}{ii}/\ion{He}{i} is always much smaller than 1 and depends
strongly on the recombination processes and hence on the densities
in the disk model. Since we use here the solar He abundance
of $\log \epsilon_{\rm He}=0.1$, the destructive reaction He$^+$ + 
CO keeps the C$^+$ abundance up throughout the disk in all models.

\subsection{The CO-to-H$_2$ ratio}

Fig.~\ref{CO-to-H2} illustrates how the The CO-to-H$_2$ ratio
changes throughout the disk for various disk models. With
increasing disk mass, the CO abundance in the disk rises
and hence the CO-to-H$_2$ ratio increases. In general, the
difference between a model with chromospheric radiation field
and one with interstellar radiation field amounts at
maximum to $\Delta$ CO-to-H$_2$ ratio $\sim 0.5$~dex.

   \begin{figure*}
   \includegraphics[width=18cm]{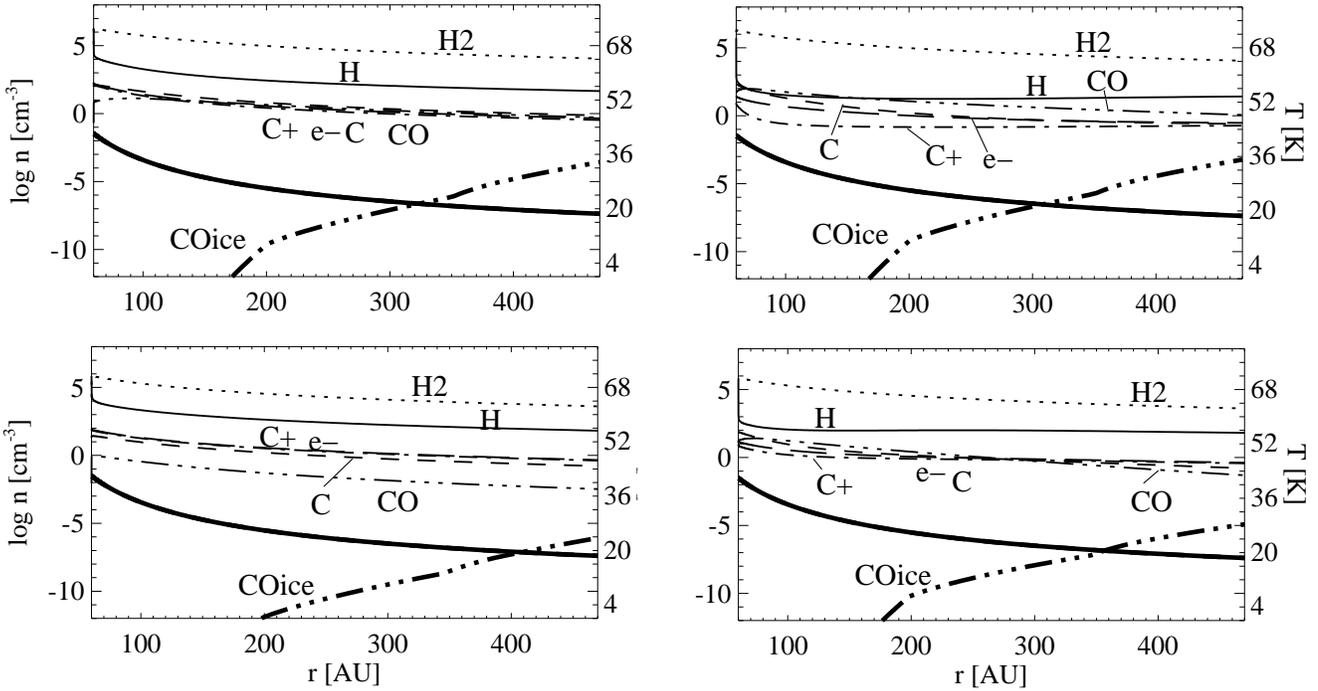}
   \caption{Densities of different species (left scale) and 
            temperature (right scale) in 2 different 
            disk models:
            (a) m3.3\_60\_CH model at the disk midplane, 
            (b) m3.3\_60\_IS model at the disk midplane, 
            (c) m3.3\_60\_CH model at one scaleheight,
            (d) m3.3\_60\_IS model at one scaleheight. The
            adopted linestyles are: H:solid, H$_2$: dotted,
            C: dashed, C$^+$: dash-dotted, CO: dash-dot-dotted,
            CO$_{\rm ice}$: dash-dot-dotted (thick), e$^-$:
            long dashed, $T$: solid (thick).}
    \label{m3.3_chrom_m3.3_ISH_chem}
    \end{figure*}

For an interpretation of observations, which give rather an integrated 
picture of the CO in the disk, it is better to use the integrated
[CO/H$_2$] ratio of the various disk models (see Table~\ref{tab:CO-to-H2}) 
\begin{equation}
\left[ {\rm CO/H_2} \right] = \frac{\int \epsilon({\rm CO})\,dV}{\int \epsilon({\rm H_2})\,dV}\,\,\,.
\end{equation}

   \begin{table}
      \caption[]{[CO/H$_2$] ratio for a sample of disk models.}
         \label{tab:CO-to-H2}
   \begin{tabular}{lrr}
   name             & [CO/H2] \\[2mm]
   \hline\\
   m0.33\_190\_CH  & $3.58\,10^{-7}$ \\
   m3.3\_190\_CH   & $1.12\,10^{-5}$ \\
   m0.33\_190\_IS  & $1.89\,10^{-6}$ \\ 
   m3.3\_190\_IS   & $6.22\,10^{-5}$ \\
   m3.3\_60\_CH    & $7.53\,10^{-5}$ \\
   m3.3\_60\_IS    & $3.65\,10^{-4}$ \\
   \hline\\
   \end{tabular}
   \end{table}

Given the tenuous character of these disks, an IS UV radiation 
field would penetrate the disk from all directions and therefore lead to
a much higher [H$_2$/CO] ratio even for the 3.3~M$_{\oplus}$ model. 
Thus, we draw the tentative conclusion that the [CO/H$_2$] conversion
factor in these disks can be orders of magnitude smaller than the canonical
value for molecular clouds, [CO/H$_2$]$\sim 10^{-4}$. These disks resemble
in their nature generally the photon dominated regions (PDR) on the
surfaces of molecular clouds, very high UV irradiation and densities
ranging from $10^3$ to a few times $10^5$~cm$^{-3}$.

\subsection{The role of dust}

As already stated above, dust shielding plays no role for
the photochemistry in the disk. On the other hand, the
dust grain surfaces are the location where molecular hydrogen
is formed. Hence, the H$_2$ formation rate
\begin{equation}
R_{\rm gr} = 3\times 10^{-17} \left(\frac{T}{\rm 100 K} \right)^{1/2} 
             \frac{\sigma_{uv}}{6 \times 10^{-22} {\rm cm^2}}~~~{\rm cm^3 s^{-1}} 
\end{equation}
is higher in 
models with a higher dust-to-gas mass ratio, e.g.\
m0.033\_190\_CH as compared to m3.3\_190\_CH. $\sigma_{uv}$
denotes here the effective UV attenuation cross section per
H nucleus. Assuming a low dust-to-gas mass ratio of $1.73\,10^{-2}$ for
the 0.033~M$_{\oplus}$ model prevents formation of H$_2$ 
throughout the whole disk.  

The alternative model (Sylvester \& Skinner \cite{SylvesterSkinner:1996})
with an inner radius of 60~AU and a dust mass of 0.77~$M_\oplus$
has an even higher dust-to-gas mass ratio (m3.3\_60\_CH: 0.23 compared
to m3.3\_190\_CH: $1.73\,10^{-2}$) and therefore H$_2$ extends up to two 
scaleheights
and CO up to one scaleheight in the disk. Additionally, the
freezing out of CO is slightly enhanced, but still not efficient
enough to obtain a high CO ice abundance. But even in these
disk models with a high dust mass, dust absorption is not a
dominant shielding process.

\subsection{The role of the chromosphere}

The outer parts of the disk beyond 400~AU are clearly dominated
by the interstellar radiation field. Nevertheless, for the
inner gas disk the chromospheric UV radiation is the driving
mechanism in chemistry. If one neglects the presence of a
chromosphere, the resulting H and CO abundances differ
by more than a factor 10, especially in regions where most of
the disk mass is located (see Fig.~\ref{m3.3_chrom_m3.3_ISH_chem}).

For somewhat younger stars, the chromosphere will be even more 
important, because the stronger stellar activity will give rise to
an even higher UV photon flux. At a certain point back in time, the disk has
been optically thick and actively accreting; therefore UV photons 
and X-rays from the accretion process have probably determined the 
chemistry and temperature balance in the upper layers of the
protoplanetary disk. At which stellar age the transition 
to optically thin disks occurs as well as the point in time when 
accretion ceases and stellar activity takes over has still to be 
determined.

\section{Conclusion}

In this paper, we investigate the chemical structure of tenuous circumstellar
disks around young solar-type stars using the disk models developed
by Kamp \& Bertoldi (\cite{KampBertoldi:2000}). Previous observational
studies have neglected the possible influence of a chromosphere for 
solar-type stars. Especially in an earlier evolutionary stage (WTTS), the UV
radiation from the chromosphere is much stronger than that of the present
Sun and may determine the chemistry in the circumstellar surrounding.

The main results of this work are:
\begin{itemize}
\item We present a recipe to calculate the chromospheric UV radiation 
      field of young solar-type stars as a function of age.
\item The CO-to-H$_2$ ratio in low mass disks around solar-type stars
      is smaller than the canonic value of $10^{-4}$ for molecular clouds
      and resembles more the values found for the PDR surface
      layers of molecular clouds.
\item The dust-to-gas mass ratio has a strong influence on the 
      molecular hydrogen abundance in disks. Low mass disks, in which 
      hydrogen would otherwise be atomic, can become molecular in the
      presence of a high dust-to-gas mass ratio. Note that since these
      disk models are optically thin in the UV, this is not due to dust 
      shielding, but due to an enhanced H$_2$-formation rate.
\item The chromosphere of a 70~Myr old star dissociates the CO in the
      inner disk ($r < 400$~AU) of a solar-type star. Due to self-shielding 
      effects, H$_2$ is much less affected than CO. In the outer disk,
      the stellar UV radiation field is smaller than the typical interstellar
      UV field.
\end{itemize}

This work presents a first step towards the modeling of protoplanetary
disks around young solar-type stars. The model is for example limited by
the fact that we did not include an interstellar UV radiation field in
a self-consistent way (2D radiative transfer). Moreover, at least a 
simple line radiative transfer has to be included in order to (a) be able 
to calculate the gas temperature in these disks self-consistently and 
(b) to extend these disk models to higher masses. 
%

\begin{acknowledgements}
The authors thank Ewine van Dishoeck for many fruitful discussions 
during the course of this work. We are grateful to the referee, whose
comments helped to make this paper clearer and more concise and to
Jeff Valenti for a careful reading of the manuscript. 
Astrochemistry in Leiden is supported by a SPINOZA grant from the Netherlands
Organization for Scientific Research (NWO).
\end{acknowledgements}


\begin{thebibliography}{}

  \bibitem[1997]{Ayres:1997} Ayres T.R. 1997, JGR Planets 102, 1641

  \bibitem[1998]{Coulson_etal:1998} Coulson I.M., Walther D.M., Dent W.R.F. 1998, MNRAS 296, 934

  \bibitem[1994]{DorrenGuinan:94} Dorren, J.D., Guinan, E.F. 1994, ApJ 428, 805

  \bibitem[1997a]{Dunkin_etal:1997a} Dunkin S.K., Barlow M.J., Ryan S.G. 1997, MNRAS 286, 604

  \bibitem[1997b]{Dunkin_etal:1997b} Dunkin S.K., Barlow M.J., Ryan S.G. 1997, MNRAS 290, 165

  \bibitem[2000]{Greaves_etal:2000} Greaves J.S., Coulson I.M., Holland W.S. 2000, MNRAS 312, L1

  \bibitem[1968]{Habing:1968} Habing H.J. 1968, Bull.\ Astr.\ Inst.\ Netherlands 19, 421

  \bibitem[1999]{JuraKahane:1999} Jura M., Kahane C. 1999, ApJ 521, 302

  \bibitem[2002]{Kalas_etal:2002} Kalas P., Graham J.R., Beckwith S.V.W., Jewitt D.C., Lloyd J.P.
   2002, ApJ 567, 999

  \bibitem[2000]{KampBertoldi:2000} Kamp, I., \& Bertoldi, F. 2000, A\&A, 353, 276

  \bibitem[1992]{Kurucz:1992} Kurucz R.L. 1992, Rev.\ Mex.\ Astron.\ Astrofis.\ 23, 181

  \bibitem[2001]{Mora_etal:2001} Mora A., Mer\'{i}n B., Solano E., Montesinos B., de Winter D., et al.
   2001, A\&A 378, 116

  \bibitem[1986]{Scoville_etal:1986} Scoville N.Z., Sargent A.I., Sanders D.B., Claussen M.J.,
   Masson C.R., Lo K.Y., Phillips T.G. 1986, ApJ 303, 416

  \bibitem[1996]{SylvesterSkinner:1996} Sylvester R.J., Skinner C.J. 1996, MNRAS 283, 457

  \bibitem[1996]{Sylvester_etal:1996} Sylvester R.J., Skinner C.J., Barlow M.J., Mannings V. 
   1996, MNRAS 279, 915

  \bibitem[2001]{Zuckerman:2001} Zuckerman B. 2001, ARAA 39, 549

  \bibitem[1995]{Zuckerman_etal:1995} Zuckerman B., Forveille T., Kastner J.H. 1995, Nature 373, 494

\end{thebibliography}
\end{document}